\documentclass[12pt]{article}
\usepackage[top=1.5cm, bottom=1.5cm, left=1.5cm, right=1.5cm]{geometry}
\usepackage{longtable}
\usepackage{natbib}
\usepackage{amsmath}
\usepackage{txfonts}
\usepackage{graphicx}
\usepackage{amssymb}
\usepackage[labelfont=bf]{caption}
\begin{document}
\title{A cyclical period variation detected in the updated orbital period analysis of TV Columbae}
\author{Zhibin Dai $^{1,2}$, Shengbang Qian $^{1,2}$, Eduardo Fern\'{a}ndez Laj\'{u}s $^{3}$ and G. L. Baume $^{3}$}

\footnotetext[1]{\scriptsize{National Astronomical Observatories/Yunnan
Observatory, Chinese Academy of Sciences, P. O. Box 110, 650011
Kunming, P. R. China.}}

\footnotetext[2]{\scriptsize{Key Laboratory for the Structure and Evolution of Celestial Objects, Chinese Academy of Sciences, P. R. China.}}

\footnotetext[3]{\scriptsize{Facultad de Ciencias Astron\'{o}micas y
Geof\'{i}sicas, Universidad Nacional de La Plata Paseo del Bosque
s/n, 1900, La Plata, Pcia. Bs. As., Argentina.}}

\maketitle

\begin{abstract}
\small

\noindent{The two CCD photometries of the intermediate polar TV Columbae are made for obtaining the two updated eclipse timings with high precision. There is an interval time $\sim17yr$ since the last
mid-eclipse time observed in 1991. Thus, the new mid-eclipse times
can offer an opportunity to check the previous orbital ephemerides. A calculation indicates that the orbital ephemeris derived by
\cite{aug94} should be corrected. Based on the proper linear
ephemeris \citep{hel93}, the new orbital period analysis suggests a
cyclical period variation in the O-C diagram of TV Columbae. Using Applegate's mechanism to explain the periodic
oscillation in O-C diagram, the required energy is larger than that a M0-type star can afford over a complete variation period
$\sim31.0(\pm3.0)yr$. Thus, the light travel-time effect indicates
that the tertiary component in TV Columbae may be a
dwarf with a low mass, which is near the mass lower limit
$\sim0.08M_{\odot}$ as long as the inclination of the third body high enough.}

\end{abstract}

\begin{bfseries}
\noindent{Stars: cataclysmic variables; Stars : binaries : eclipsing; Stars : individual (TV Columbae)}
\end{bfseries}


\section{Introduction}

An intermediate polar TV Columbae was first discovered
as a hard X-ray source 2A 0526-328 by \cite{coo78}. The subsequent
photometries and spectroscopies detected five periods of 32 min, 5.2
hr, 5.5 hr, 6.3 hr and $\sim$4 day, which corresponding to the spin
modulation in X-ray \citep{sch85}, the beat period between 5.5 hr and 4 day \citep{mot81}, the binary orbital period \citep{hut81}, the
permanent positive superhump period \citep{ret03} and the nodal
precession of the accretion disc period \citep{hel93}, respectively.
No until \cite{hel91} was the eclipse in TV Columbae
reported. Except for the 32 min detected in X-ray observation, the
other four periods presented in optical regime give rise to the high
variabilities in eclipse light curves of TV Columbae
\citep{hel91,hel93,aug94}. Although the eclipse light curves of
TV Columbae show cycle-to-cycle variations, a shallow
eclipse of part of the accretion disc is established. \cite{hel93}
and \cite{aug94} have pointed out that the eclipse of TV Columbae in a certain case (maybe outburst) is clearly modulated by
a broader quasi-sinusoidal modulation. Moreover, \cite{hel93} found a peculiar phenomenon in TV Columbae that its eclipse can
totally disappear in the light curve observed in January 1st, 1986.

Since plenty of investigators focused on the variations of 5.2 hr and 4 day periods, few analyses in its orbital period were made in the
past. However, many 5.2 hr period analyses indicate that this period
is not stable and can change non-monotonically with time
\citep{aug94}. In addition, it is difficult to detect the variations
of 4-d modulation because of the short observation time in most case. Thus, the stable and distinct photometric period of 5.5 hr in
TV Columbae can provide a best opportunity to probe the
possible evolutional state and inner interaction between two
components. The first orbital period analysis \citep{hel93} suggested that the quadratic term in the O-C diagram was suspected. Then
\cite{aug94} and \cite{ran04} never detected the changes in its
orbital period. A time span of about 17 years, from the last eclipse
timing \citep{hel93} to now, implies that an updated and available
orbital period analysis is expected to detect the possible variations in orbital period of TV Columbae.

In this paper, the two light curves near the mid-eclipse by CCD
photometries and two new eclipse timings of TV Columbae
with high precision are presented in Sect. 2. Then Sect. 3 deals with the details of the updated O-C analysis in the orbital period of 5.5
hr. Finally, the probable discussions for the observed orbital period changes are made in Sect. 4 and our principal conclusions in Sect. 5.

\section{Observations}

Two new times of light minimum are obtained from our CCD photometric
observations with the VersArray 1300B CCD camera attached to the
2.15-m Jorge Sahade telescope at Complejo Astronomico El Leoncito
(CASLEO), San Juan, Argentina. Both photometries of TV Columbae in V-filter were carried out on November 18 and 20, 2009.
Two nearby stars which have the similar brightness in the same
viewing field of telescope are chosen as the comparison star and the
check star, respectively. All images were reduced by using PHOT
(measure magnitudes for a list of stars) of the aperture photometry
package of IRAF.

Two partial eclipse light curves shown in Fig. 1 indicate obviously
night-to-night variations, which present different eclipse depth
($0^{m}.2\sim0^{m}.45$) at the different cycle. A nearly flat-toped
maximum with a phase range $0.6\sim0.9$ is detected in the eclipse
light curve observed in November 18, 2009, which is never found in
the previous photometries \citep{hel91,hel93,aug94}. Moreover, we
obtained a distinct eclipse profile with a long egress in the other
observation in November 20, 2009. The orbital light curves of
TV Columbae in outburst are similar as the sinusoidal
curves (see the photometries in 1987, 1988 and 1991 outbursts), which are totally different with those observed in quiescence. Thus,
considering the extreme changes in the eclipse profiles at different
epochs, the eclipse timings with high precision are important for the investigation of variations in orbital period of TV Columbae. Two accurate times of mid-eclipse were derived by using a
parabolic fitting method to the very deepest part of the eclipse.
Including the previous 43 times of light minimum for TV Columbae \citep{hel91,hel93,aug94}, we listed all 45 available times of light minimum covering near 30 yr in Table 1.

\begin{figure}
\centering
\includegraphics[width=9.0cm]{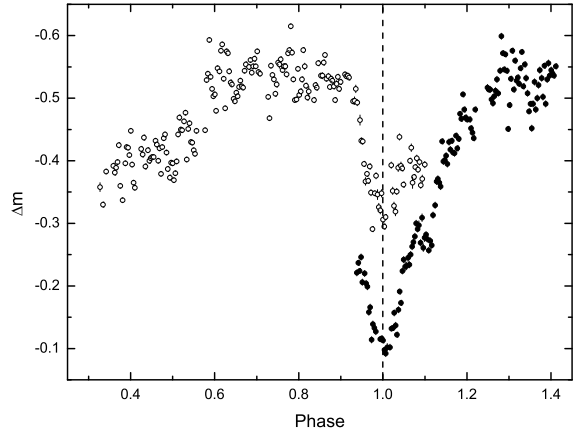}
\caption{The eclipse parts of light curves of intermediate polar TV
Columbae in V-filter measured on 2009 November 18 and 20 by using
2.15-m Jorge Sahade telescope at CASLEO, are plotted in open and
solid circles, respectively. The dash line indicates the
mid-eclipse.}\label{Fig. 1}
\end{figure}

\begin{figure}
\centering
\includegraphics[width=9.0cm]{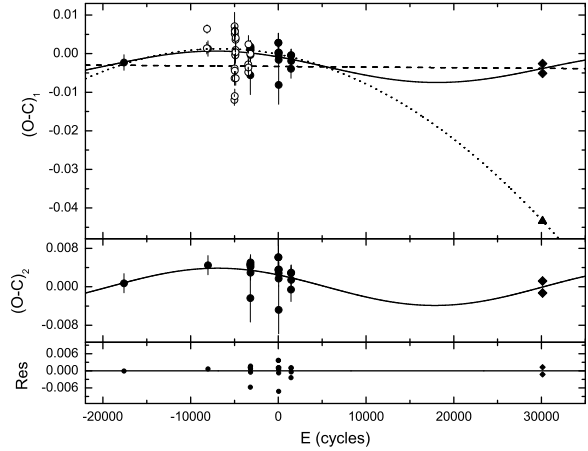}
\caption{The $(O-C)_{1}$ values of TV Columbae are calculated by Eq.
2. The best fit linear and sinusoidal curves are plotted by dash line and solid line, respectively. The two $(O-C)_{1}$ values at the same
cycles as our data points, which are predicted by the quadratic
ephemeris \citep{hel93} denoted by the dotted line, are plotted by
the solid upper triangles. The open circles represent the data points from \cite{aug94}, the filled diamonds and circles the two data
points we obtained and the data from the other papers, respectively.
After removing the linear element from the $(O-C)_{1}$ diagram and
the data from \cite{aug94}, the $(O-C)_{2}$ values shown in the
middle panel suggest a significant cyclical period change. The
residuals and their linear fitted solid line are presented in the
bottom panel. The ordinates of all three panels are in day
units.}\label{Fig. 2}
\end{figure}

\begin{figure}
\centering
\includegraphics[width=9.0cm]{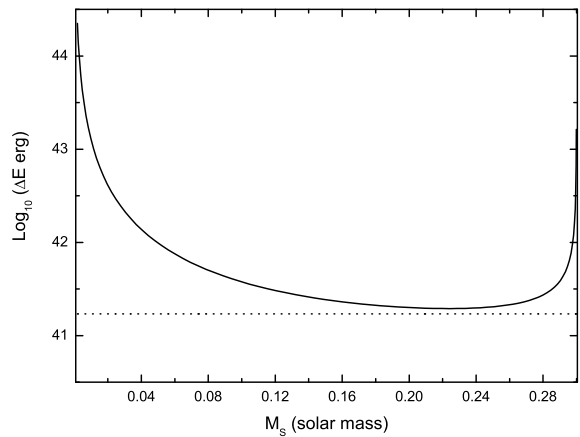}
\caption{The solid line denotes the energy required by Applegate's
mechanism corresponding to the different shell mass. The dash line
refers to the total radiant energy of a M0-type star over the whole
oscillation period $\sim31yr$.}\label{Fig. 3}
\end{figure}

\begin{figure}
\centering
\includegraphics[width=9.0cm]{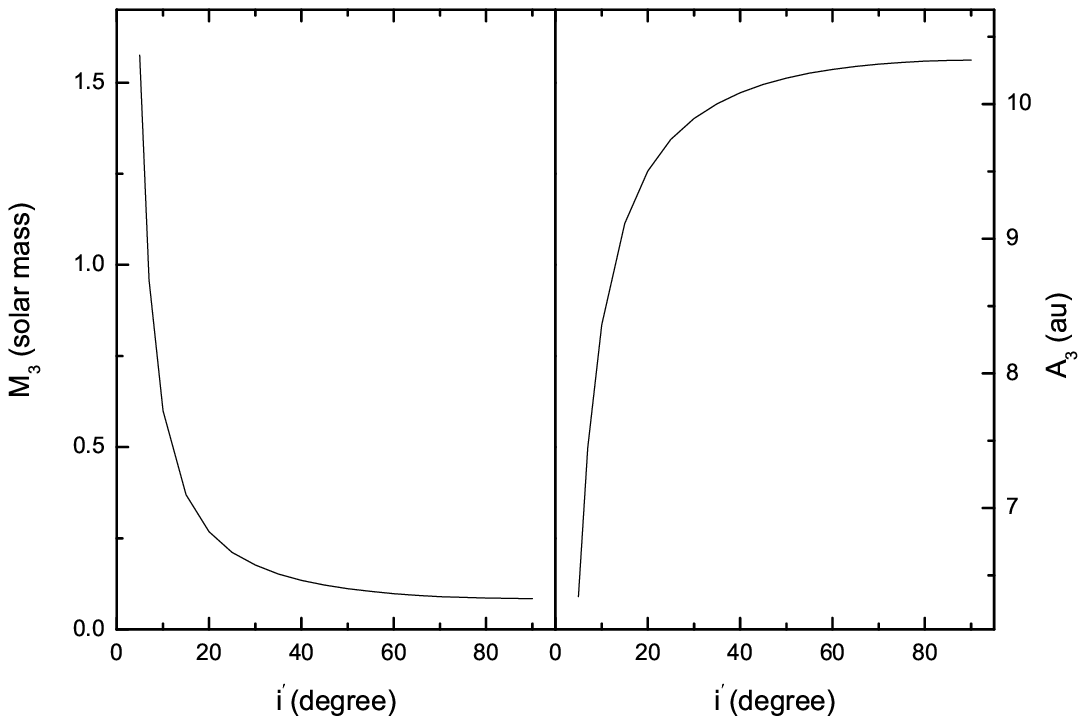}
\caption{The masses and separations of the tertiary component in TV
Columbae vs. its orbital inclination $i^{'}$ are plotted in the left
and right panels, respectively.}\label{Fig. 4}
\end{figure}

\section{Analysis of orbital period change}

The data from AAVSO an International Database
(http://www.aavso.org/data/download/) suggest that TV Columbae is not in outburst during our observations, and from Fig. 1 it cannot be seen that a dip around orbital phase $\sim1.2$, which is different from the light curves in 1987 I and 1988 obtained by
\cite{aug94}. A recent orbital ephemeris \citep{aug94,ran04},
\begin{equation}
T_{min}=HJD\;2447151.2324(11)+0^{d}.22859884(77)E,
\end{equation}
is used to check whether the updated two eclipse timings obey the
regular 5.5 hr orbital modulation. The two cycles 35007.226 and
35015.214 corresponding to the observed mid-eclipse times HJD
2455153.843579(76) and HJD 2455155.669797(71), respectively, are
calculated, which means that both observed orbital eclipse minima are about $0.2$ in phase later than the predicted times. This large
discrepancy implies that TV Columbae should seriously
expand its orbit in an extreme short timescale ($\sim30yr$), which is obviously not supported by observations. This means that either the
two derived eclipse timings or the orbital ephemeris Eq. 1 are wrong. However, the orbital ephemeris derived by \cite{hel93},
\begin{equation}
T_{min}=HJD\;2448267.4895(7)+0^{d}.22860034(16)E,
\end{equation}
can accurately predict the mid-eclipse times are HJD 2455153.846142
and HJD 2455155.674945, which corresponds to the O-C values
$-0^{d}.0026$ and $-0^{d}.0051$, respectively. Therefore, although
both light curves we obtained are not complete and the multi-periodic modulations badly affect the profiles of orbital light curves, we can ensure that the derived mid-eclipse times are correct. In order to
find the error in Eq. 1, we carefully compared both ephemerides, and
then found that the small discrepancy $\sim1^{d}.5\times10^{-6}$ in
the average orbital period between Eq. 1 and Eq. 2 is accumulated to
$\sim0^{d}.053$ (i.e. $\sim0.2$ in phase) over 30 years. Moreover,
the error in average orbital period may explain the deviations and
phase shifts of eclipse times in \cite{aug94}.

The O-C values of 45 eclipse timings for TV Columbae
listed in column 6 of Table 1 are calculated by using Eq. 2. The new
O-C diagram shown in Fig. 2 suggests a possible cyclical variation
instead of the secular decrease derived by \cite{hel93}, because
Hellier's quadratic ephemeris predicted the O-C values at the same
cycles as our data are about $-0^{d}.0434$ shown in the upper panel
of Fig. 2, which are clearly distinct with our observations.
Therefore,we attempted to use a linear-plus-sinusoidal ephemeris to
fit the O-C values. Since the large scatters from $\sim-10000$ cycles to $\sim0$ cycles obviously exceed the errors of the mid-eclipse
times, the probable weights should be reconsidered instead of those
calculated from the uncertainties of the mid-eclipse times. The
incorrect orbital ephemeris and the possible deviations in eclipse
times derived by \cite{aug94} imply that the weights of their data
should be lower than the others. Therefore, a least-square solution
for the O-C values of TV Columbae leads to
\begin{equation}
\begin{aligned}
(O-C)_{1} &
=-3^{d}.32(\pm0^{d}.07)\times10^{-3}-1^{d}.63(\pm0^{d}.28)\times10^{-8}E\\
&+3^{d}.87(\pm0^{d}.09)\times10^{-3}\sin[0^{\circ}.0073(\pm0^{\circ}.0007)E+140^{\circ}.4(\pm1^{\circ}.2)].
\end{aligned}
\end{equation}
Although the early data present large scatters, all the
high-precision data except for the two data points with large error
bars $0^{d}.005$ \citep{hel93} well support a possible cyclical
variation trend in O-C diagram after removing the data from
\cite{aug94} and the linear element, which is clearly shown in the
middle panel of Fig. 2. Moreover, the fitting $\chi^{2}\sim1.2$
suggests that a cyclical period variation with a period of
$31.0(\pm3.0)yr$ is significant.

\section{Discussion}

The radial velocity measurements \citep{hut81,hel93} and the
spectroscopic analyses \citep{vrt96,cro99,ish99,ram00} never obtained the consistent system parameters of TV Columbae.
According to the classification of TV Columbae as a
permanent superhump system \citep{ret03}, TV Columbae
may be an extreme system with a mass ratio $\sim0.3$, which is in
contradiction with that calculated by the relation between the
orbital period and the secondary mass in CVs \citep{smi98}. Moreover, assuming that the white dwarf in TV Columbae has a
typical mass of $\sim1M_{\odot}$, the secondary mass can be estimated to be $\sim0.3M_{\odot}$, which may be agree with the estimated
spectra type of the secondary late K or early M type
\citep{beu00,smi98}. Thus, we adopted a combined mass of
$1M_{\odot}+0.3M_{\odot}$ for the eclipsing pair of TV Columbae.

Considering that the eclipse of the intermediate polar TV Columbae may be of a partial accretion disc caused by the magnetic of the white dwarf, the changes in the partial disc may produce the changes in the O-C diagram shown in Fig. 2. Moreover, a cyclical period variation can be interpreted by two plausible
mechanisms, which are Applegate's mechanism \citep{app92} and light
travel-time effect. As for the former, a careful calculation for the
variation of quadrupole momentum suggests that $\triangle
Q=-2.03(\pm0.05)\times10^{48}g\;cm^{2}$. Considering the secondary in TV Columbae is known little, the energies required to
cause the observed O-C oscillation corresponding to different assumed shell masses are calculated. The lowest required energy $\triangle
E_{min}\sim1.9\times10^{41}erg$ at $M_{s}\simeq0.22M_{\odot}$, is
larger than the total radiant energy of a M0 type star over a
complete variation period $\sim31yr$, $E_{0}\sim1.7\times10^{41}erg$.
Therefore, a solar-type magnetic cycle may have difficulty explaining the observed cyclical period variation in TV Columbae as the previous analysis for other CVs \citep{dai09,dai10}.

Another plausible mechanism for explaining the cyclical period
variation in O-C diagram is the light travel-time effect caused by a
perturbations from a tertiary component. If the two data points at
cycles -3193 and 26 with large error bars can be omitted, then the
cyclical period changes shown in the middle panel of Fig. 2 would
well support an excellent sinusoidal fit. By using the amplitude of
sinusoidal curve and the Third Kepler Law, the projected distance
$a^{'}\sin(i^{'})$ from binary to the mass center of the triple
system and the mass function of the third component $f(m_{3})$ can be calculated to be $0.67(\pm0.02)au$ and
$3.1(\pm0.2)\times10^{-4}M_{\odot}$, respectively. According to the
both relationships described in Fig. 4, the mass of the third star in TV Columbae is very close to the mass lower limit of a
star with stable hydrogen burning ($\sim0.08M_{\odot}$) when the
inclination of the third star is higher enough. In addition, the
distance from the third body to the mass center of system is
$\sim10.3 au$, which is over two orders of magnitude larger than the
separation of binary. Thus, this third star can survive the previous
common envelope evolution of the parent binary.

\section{Conclusion}

Two new light curves near mid-eclipse with obviously night-to-night
variations are obtained. Moreover, a near flat-top shape maximum in
the light curve observed in November 18, 2009 with a phase range
$0.6\sim0.9$ is first detected. Both updated eclipse timings with
high precision for intermediate polar TV Columbae we
observed, which indicate a phase shift of $\sim0.2$ by using Eq. 1,
can be accurately predicted by Eq. 2. A compare between the two
orbital ephemerides indicates that the average orbital period derived by \cite{aug94} is incorrect and the errors are accumulated to
$\sim0^{d}.053$ (i.e. $\sim0.2$ in phase) over the 30 years. This
means the Augusteijn's orbital ephemeris is not appropriate for
calculating the orbital phase of TV Columbae. Based on
the probable linear ephemeris, a detailed orbital period analysis
suggests a significant cyclical period variation with a period of
$31.0(\pm3.0)yr$ in its O-C diagram, which can not be explained by
Applegate's mechanism. Accordingly, a light travel-time effect is
applied to interpret the cyclical period change. The third body in
TV Columbae may be a low mass dwarf as long as the
orbital inclination of tertiary component high enough. Since TV Columbae is a complex system and the explanations for its updated O-C diagram is only based on the two new mid-eclipse timings, the cyclical period change shown in Fig. 2 should be checked by the more data in the future. Although the system parameters of TV Columbae are ambiguous, which means that any conclusive result cannot be obtained, the discussions based on some plausible assumptions are still important to the further understanding for TV Columbae. Therefore, the more observations including CCD photometries with a longer base line and spectroscopies with high resolution are needed to probe such multi-periodic system.

\section*{Acknowledgements}

\small{This work was partly Supported by Special Foundation of President of The Chinese Academy of Sciences and West Light Foundation of The Chinese Academy of Sciences, Yunnan Natural Science Foundation (2008CD157), the Yunnan Natural Science Foundation (No. 2005A0059M) and Chinese Natural Science Foundation (No.10573032, No. 10573013 and No.10433030). CCD photometric observations of TV Columbae were obtained with the 2.15-m Jorge Sahade telescope at CASLEO, San Juan, Argentina. We thank the referee very much for the helpful comments and suggestions that helped to improve this paper greatly.}


\begin{center}
\begin{longtable}{p{3cm}ccccrc}
\caption{The 45 eclipse timings for the intermediate polar TV Columbae.}\\
\hline\hline
\hspace{2em}JD.Hel. & type & error & Method & E (cycle) & $(O-C)_{1}^{d}$ & Ref.\\
\hspace{1.5em}2400000+&&&&&&\\
\hline
\endfirsthead
\caption{Continued.}\\
\hline\hline
\hspace{2em}JD.Hel. & type & error & Method & E (cycle) & $(O-C)_{1}^{d}$ & Ref.\\
\hspace{1.5em}2400000+&&&&&&\\
\hline
\endhead
\hline
\endfoot
\hline \multicolumn{7}{p{14cm}}{\scriptsize{References: (1)
\cite{mot81}; (2) \cite{aug94}; (3) \cite{hel93}; (4) \cite{hel91};
(5) This paper.}}
\endlastfoot
44243.664000 & pri & .002    & pe  & -17602 & -.0023   & (1)\\
46403.712300 & pri & .0015   & pe  & -8153  &  .0014   & (2)\\
46409.660900 & pri & .0011   & pe  & -8127  &  .0064   & (2)\\
46431.372900 & pri & .002    & ccd & -8032  &  .0013   & (3)\\
47122.647200 & pri & .0014   & pe  & -5008  & -.012    & (2)\\
47124.712600 & pri & .0036   & pe  & -4999  & -.0038   & (2)\\
47125.627000 & pri & .0015   & pe  & -4995  & -.0038   & (2)\\
47126.767400 & pri & .0018   & pe  & -4990  & -.0064   & (2)\\
47127.695300 & pri & .0036   & pe  & -4986  &  .0071   & (2)\\
47128.607700 & pri & .0019   & pe  & -4982  &  .0051   & (2)\\
47129.734400 & pri & .0018   & pe  & -4977  & -.011    & (2)\\
47130.655600 & pri & .0009   & pe  & -4973  & -.0044   & (2)\\
47131.808900 & pri & .0015   & pe  & -4968  &  .0059   & (2)\\
47134.780500 & pri & .002    & pe  & -4955  &  .0057   & (2)\\
47146.661800 & pri & .002    & pe  & -4903  & -.00023  & (2)\\
47147.804600 & pri & .0021   & pe  & -4898  & -.00043  & (2)\\
47148.723000 & pri & .0032   & pe  & -4894  &  .0036   & (2)\\
47150.777800 & pri & .002    & pe  & -4885  &  .00096  & (2)\\
47151.691600 & pri & .0022   & pe  & -4881  &  .00036  & (2)\\
47153.742200 & pri & .0018   & pe  & -4872  & -.0064   & (2)\\
47155.581500 & pri & .0021   & pe  & -4864  &  .0041   & (2)\\
47481.785300 & pri & .0013   & pe  & -3437  & -.0048   & (2)\\
47482.701700 & pri & .002    & pe  & -3433  & -.0028   & (2)\\
47486.815700 & pri & .0021   & pe  & -3415  & -.0036   & (2)\\
47487.736100 & pri & .0023   & pe  & -3411  &  .0024   & (2)\\
47537.111400 & pri & .0016   & ccd & -3195  & -.000014 & (4)\\
47537.563000 & pri & .005    & ccd & -3193  & -.0056   & (3)\\
47538.484400 & pri & .0016   & ccd & -3189  &  .0014   & (3)\\
47539.398700 & pri & .0016   & ccd & -3185  &  .0013   & (3)\\
47540.540100 & pri & .0016   & ccd & -3180  & -.00032  & (3)\\
47541.456600 & pri & .0016   & ccd & -3176  &  .0018   & (3)\\
47542.370100 & pri & .0016   & ccd & -3172  &  .00088  & (3)\\
48266.349300 & pri & .0025   & ccd & -5     &  .0028   & (3)\\
48267.489700 & pri & .0016   & ccd & 0      &  .0002   & (3)\\
48268.406700 & pri & .0025   & ccd & 4      &  .0028   & (3)\\
48273.425000 & pri & .005    & ccd & 26     & -.0081   & (3)\\
48274.345900 & pri & .0016   & ccd & 30     & -.0016   & (3)\\
48275.490000 & pri & .0016   & ccd & 35     & -.00051  & (3)\\
48278.462600 & pri & .0016   & ccd & 48     &  .00028  & (3)\\
48595.301900 & pri & .0016   & ccd & 1434   & -.00049  & (3)\\
48599.413300 & pri & .0025   & ccd & 1452   & -.0039   & (3)\\
48602.388600 & pri & .0016   & ccd & 1465   & -.0004   & (3)\\
48603.530100 & pri & .0025   & ccd & 1470   & -.0019   & (3)\\
55153.843579 & pri & .000076 & ccd & 30124  & -.0026   & (5)\\
55155.669797 & pri & .000071 & ccd & 30132  & -.0051   & (5)\\
\end{longtable}
\end{center}

\end{document}